\documentclass[twocolumn,showpacs,amsmath,amssymb,prb,graphics,graphicx]
{revtex4-2}
     \usepackage{graphicx}
\usepackage{bm}
\usepackage{dcolumn}
\begin{document}

\title{Dissipation of moving  vortices in thin films}

\author{ V.G. Kogan }
\affiliation{ Ames Laboratory-DOE, Ames, Iowa 50011 }
  \author{N. Nakagawa}
 \affiliation{Iowa State University, Ames, IA 50011, USA}   

     \date{\today}
     
\begin{abstract}
Moving vortices in thin superconducting films are considered within the time-dependent London description. The dissipation due to out-of-core normal excitations for two vortices moving together turns out to have a minimum for the separation vector $\bm a$ parallel to the velocity  and equal to $a_m \approx 2.2\, \Lambda$,  where $\Lambda $ is the Pearl length. The minimum entropy production  suggests that moving vortices should have a tendency to form chains along the velocity with a period of the order $a_m$. 
\end{abstract}


\maketitle

\section{Introduction}

 Problems of vortex dynamics in superconductors have recently come back to the community attention because new and more accurate experimental techniques become available. Vortex velocities well above the speed of sound are  now attainable along with new methods of measuring  field distributions \cite{Eli,Denis}.

 Moving vortices, pushed by the Lorentz force due to applied transport current, dissipate energy replenished by the current source. In this situation, the heat transfer should be taken into account \cite{Denis}, just to mention one of the complications.
One of the   facts attracting attention   is that moving vortices tend to form chains extended along the velocity. The chains have  periods $a>> \xi$, the vortex core size, so that the linear London approach may provide useful insights
notwithstanding the London inability to treat the vortex core physics.

In this work we study the dissipation $\cal W$ due to out-of-core quasiparticles in thin films and find that for a pair of vortices   ${\cal W}(\bm a)$ has a minimum at a finite separation $\bm a$   oriented along the pair velocity $\bm v$. The value of this separation is $a_m\approx 2.2\, \Lambda$ with the Pearl length $ \Lambda=2\lambda^2/d$ ($\lambda $ is the penetration depth of the film material and $d$ is the film thickness). According to the principle of minimum entropy production (or minimum dissipation) in stationary processes \cite{min} the system of moving vortices should have a tendency to form chains along the velocity in which vortices sit at the dissipation  minima. 
  
Within the general approach to slow relaxation processes one relates the time derivative of whatever quantity is relaxing, say $\Psi$, to the variational derivative of the   free energy functional ${\cal F}(\Psi)$, see e.g. \cite{Gork}:
 \begin{eqnarray}
- \chi \frac{\partial {\Psi}}{\partial t}= \frac{\delta  {\cal F}}{\delta {\Psi}}\,,  
  \label{1}
\end{eqnarray}
where $\chi$ is the proper relaxation time. 
The quantity of interest in our case is the vortex field distribution $\bm h(\bm r,t)$ away of the vortex core where  the London approach holds and the energy (magnetic+kinetic) is $ {\cal F} =\int d^2{\bm r}\left(h^2+\lambda^2({\rm curl} {\bm h})^2\right)/8\pi $ \cite{deGennes}: 
 \begin{eqnarray}
- \chi \frac{\partial {\bm h}}{\partial t}= \frac{\delta  {\cal F} }{\delta {\bm h}} \,. 
  \label{B1}
\end{eqnarray}
This yields
 \begin{eqnarray}
- \chi \frac{\partial {\bm h}}{\partial t}= \frac{1}{4\pi}(\bm h-\lambda^2\nabla^2\bm h)   \,, 
  \label{eqB3}
\end{eqnarray}
which reduces to the common London equation in equilibrium.

The relaxation constant $\chi$ is obtained   by comparison with the time dependent London equation \cite{TDL}, which  at   distances large relative to the core size is obtained from the assumption   that the current   consists of the normal and superconducting parts:
\begin{equation}
{\bm J}= \sigma {\bm E} -\frac{2e^2 |\Psi|^2}{mc}\, \left( {\bm
A}+\frac{\phi_0}{2\pi}{\bm
\nabla}\theta\right)  \,,
\label{current}
\end{equation}
where  $\bm A$ is the vector potential, $\Psi$ is the order parameter,  $\theta$ is the phase,  $\phi_0$ is the flux quantum, ${\bm E}$ is the electric field, and $\sigma$ is the conductivity associated with normal excitations. 
 At  these distances, $|\Psi|$ is a constant and 
acting on Eq.\,(\ref{current}) by curl one obtains \cite{TDL}:
\begin{equation}
{\bm h}- \lambda^2\nabla^2{\bm h} +\tau\,\frac{\partial {\bm h}}{\partial
t}= \phi_0  \hat{\bm z}\sum_{\nu}\delta({\bm r}-{\bm r_\nu})\,,
\label{TDL}
\end{equation}
where   ${\bm r_\nu}(t) $ is the position of the $\nu$-th vortex that may depend on time $t$, $\hat{\bm z}$ is the direction of vortices,  $\phi_0$ is the flux quantum. The relaxation time 
\begin{equation}
\tau=  4\pi\sigma\lambda^2/c^2 \,.
\label{tau}
\end{equation}
Comparing this with Eq.\,(\ref{eqB3})  one has $\chi =4\pi\tau$. In fact, the time-dependent GL equations can be obtained in a similar  manner \cite{Gork}.
 
\section{Thin films}

 Let the film of thickness $d$ be in the $xy$ plane. Integration of Eq.\,(\ref{TDL}) over the film  thickness gives for the $z$ component of the field for a Pearl vortex moving with velocity $\bm v$:
\begin{eqnarray}
\frac{2\pi\Lambda}{c}{\rm curl}_z {\bm g} + h_z  +\tau\frac{\partial h_z}{\partial t}=\phi_0 \delta(\bm r -\bm  vt)    .
\label{2D London}
\end{eqnarray}
Here, $\bm g$ is the sheet current density related to the tangential field components at the upper film face by  $2\pi\bm g/c=\hat{\bm z}\times \bm h$; $\Lambda=2\lambda^2/d$ is the Pearl length. With the help of div$\bm h=0$ this equation is transformed to:
\begin{eqnarray}
h_z -\Lambda \frac{\partial h_z}{\partial z}   +\tau\frac{\partial h_z}{\partial t}=\phi_0 \delta(\bm r -\bm  vt) .
\label{hz-eq}   
\end{eqnarray}
 
As was stressed by Pearl \cite{Pearl,deGennes},   the problem of a vortex in a thin film is
reduced to that of the stray field distribution in free space subject to the boundary condition (\ref{hz-eq}) at the film surface. Since outside the film curl$\bm h=\,\,\,$div$\bm h=0$, one can introduce a scalar potential for the {\it outside} field:
 \begin{eqnarray}
\bm h   =\bm \nabla \varphi,\qquad \nabla^2\varphi=0   \,.
\label{define _phi} 
\end{eqnarray}
The general form of the potential satisfying Laplace equation and vanishing at $z\to\infty$   is 
 \begin{eqnarray}
\varphi (\bm r, z)   =\int \frac{d^2\bm k}{4\pi^2} \varphi(\bm k) e^{i\bm k\cdot\bm r-kz}\,
\label{gen_sol} 
\end{eqnarray}
that is checked by direct differentiation.
Here, $\bm k=(k_x,k_y)$, $\bm r=( x, y)$, and  $ \varphi(\bm k)$  is the two-dimensional (2D) Fourier transform of $ \varphi(\bm r, z=0)$. In the lower half-space one has to replace $z\to -z$. 

As is done in \cite{TDL}, one applies the 2D Fourier transform to Eq.\,(\ref{hz-eq}) to obtain a linear differential equation for $h_{z\bm k}(t)$, the solution of which is:
\begin{eqnarray}
h_{z\bm k}=-k\varphi_{\bm k}  = \frac{\phi_0 e^{-i\bm k\cdot\bm v t}}{ 1+\Lambda k-i \bm k\cdot\bm v \tau }  \,.
\label{phi(k)} 
\end{eqnarray}

For two vortices separated by $\bm a$, the right-hand side of Eqs.\,(\ref{2D London}) and (\ref{hz-eq})
is 
\begin{eqnarray}
 \phi_0\left[ \delta(\bm r -\bm  vt) +   \delta(\bm r -\bm a-\bm  vt)\right],
\label{RHS}   
\end{eqnarray}
so that we obtain for the field
\begin{eqnarray}
h_{z\bm k}=  \frac{\phi_0 e^{-i\bm k\cdot\bm v t}(1+e^{-i\bm k\cdot\bm a})}{ 1+\Lambda k-i \bm k\cdot\bm v \tau }  \,.
\label{phi(k)} 
\end{eqnarray}
 
\section{Electric field and dissipation for slow motion}

This field is found from quasi-stationary Maxwell equations curl$\bm E =-\partial_t\bm h/c$ and div$\bm E =0$ \cite{LL,Gork}, which yeild in 2D Fourier space:
 \begin{equation}
   E_{x\bm k  }=-\frac{k_y}{k_x}E_{y\bm k  }= -\frac{ik_y}{ck^2} \,\frac{\partial h_{z\bm k  }}{\partial t}  \,.
\label{Es}
\end{equation}
For a pair of vortices separated by $\bm a$, we have
\begin{eqnarray}
\frac{\partial h_{z\bm k}}{\partial t} = -i \frac{\phi_0 (\bm k\cdot\bm v)\,(1+e^{-i\bm k\cdot\bm a})}{ 1+\Lambda k-i \bm k\cdot\bm v \tau } e^{-i\bm k\cdot\bm v t} \,.
\label{dh/dt} 
\end{eqnarray}
 We are interested in  motion  with constant velocity $\bm v=v\hat{\bm x}$, so that we can evaluate the fields   at $t=0$, i.e. the factor $e^{-i\bm k\cdot\bm v t}$ can be omitted.
  Then, Eqs.\,(\ref{Es}) and (\ref{dh/dt}) yield: 
 \begin{eqnarray}
     E_{x\bm k} &=&  \frac{\phi_0 v}{c} \frac{k_y k_x(1+e^{-i\bm k\bm a})}{ k^2 (1+\Lambda k)
 }  \,,\nonumber\\
    E_{y\bm k} &=& - \frac{\phi_0v }{c} \frac{ k_x^2 (1+e^{-i\bm k\bm a}) }{  k^2 (1+\Lambda k)}  \,.
\label{Exy} 
\end{eqnarray}
Since the pre-factor here contains $v$, in linear approximation in velocity the term $i \bm k\cdot\bm v \tau $ in denominators can be discarded for slow motion.

\begin{figure}[h ]
\includegraphics[width=7cm] {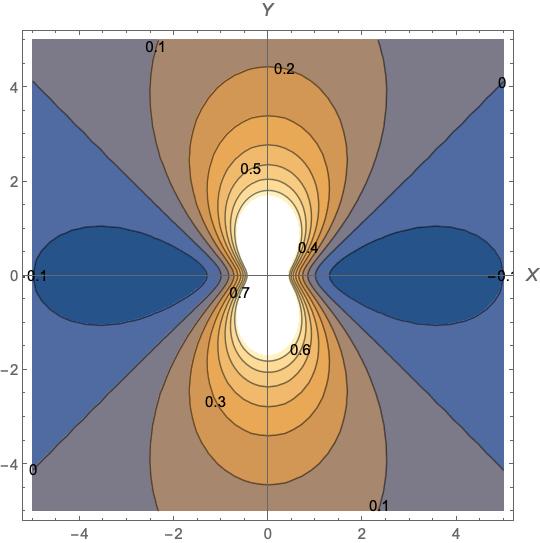}
\caption{Contours of constant dissipation $W_2(X,Y)$ for a pair of vortices, one at the origin and the other at $(X,Y)=(a_x,a_y)/\Lambda$ moving with the same velocity   along the $X$ axis.  }
\label{f1}
\end{figure}

The dissipation power  follows:
 \begin{eqnarray}
   { \cal W} &=&\sigma d\int d^2\bm r E^2 = \sigma d\int \frac{d^2\bm k}{4\pi^2} ( |E_{x\bm k}|^2+ |E_{y\bm k}|^2)\nonumber\\
   &=&\frac{\phi_0^2  v^2\sigma d}{2\pi^2c^2} \int \frac{d^2\bm k\,k_x^2(1+\cos \bm k\bm a)}
    { k^2(1+k  \Lambda  )^2}  \,.
\label{eq17} 
\end{eqnarray}
We now go to dimensionless $\bm q=\Lambda\bm k$:
 \begin{eqnarray}
   \frac{{ \cal W}}{{ \cal W}_0}  =  \int \frac{d^2\bm q \,q_x^2 (1+\cos \bm q\bm R)}{  q^2(1+  q)^2 }  =W_1+W_2 \,,
\label{calW3is} 
\end{eqnarray}
where ${\cal W}_0= \phi_0^2  v^2\sigma d/2\pi^2c^2\Lambda^2$   and $\bm R=\bm a/\Lambda$. The first contribution 
\begin{eqnarray}
  W_1=  \int \frac{d^2\bm q \,q_x^2 }  {  q^2(1+  q)^2 }=
\pi \ln\frac{1}{e\xi}
   \label{W1} 
\end{eqnarray}
where the upper limit of the divergent integral over $q$ is taken as $1/\xi$ to avoid the vortex core ($\xi$ here is the dimensionless core size). The second contribution is
 \begin{eqnarray}
&&  W_2=  \int \frac{d^2\bm q \,q_x^2 \cos \bm q\bm R}  {  q^2(1+  q)^2 } \nonumber\\
 &&   =
   \int_0^\infty \frac{d  q \,q}  { (1+  q)^2 } \int_0^{2\pi}d\phi\cos^2\phi\cos[qR\cos(\phi-\alpha)]\qquad
  \label{W2} 
\end{eqnarray}
with $\phi$ being the azimuth of $\bm q$ and 
$\alpha $ is the angle between $\bm R=\bm a/\Lambda$ and $X$.
After substitution $\beta=\phi-\alpha$, the angular integral takes the form
\begin{eqnarray}
   \int_0^{2\pi}d\beta\cos^2(\beta&+&\alpha)\cos(qR\cos\beta)\nonumber\\
  & =&2\pi \left(\frac{J_1(qR)}{qR}-J_2(qR)\cos^2\alpha\right),\qquad
  \label{in-angle} 
\end{eqnarray}
where $J_{1,2}$ are Bessel functions of the first kind. 
The integration over $q$ can be done analytically resulting in a cumbersome combination of Bessel  and Hypergeometric functions. We avoid this by doing this integration numerically. The contours of ${\cal W}_2(X,Y)=\,$const are shown in Fig.\,\ref{f1}. Note that the contours of the total dissipation $\cal W$ = const, are in fact  the same because $W_1$ is a coordinate independent constant.

\begin{figure}[h ]
\includegraphics[width=7cm] {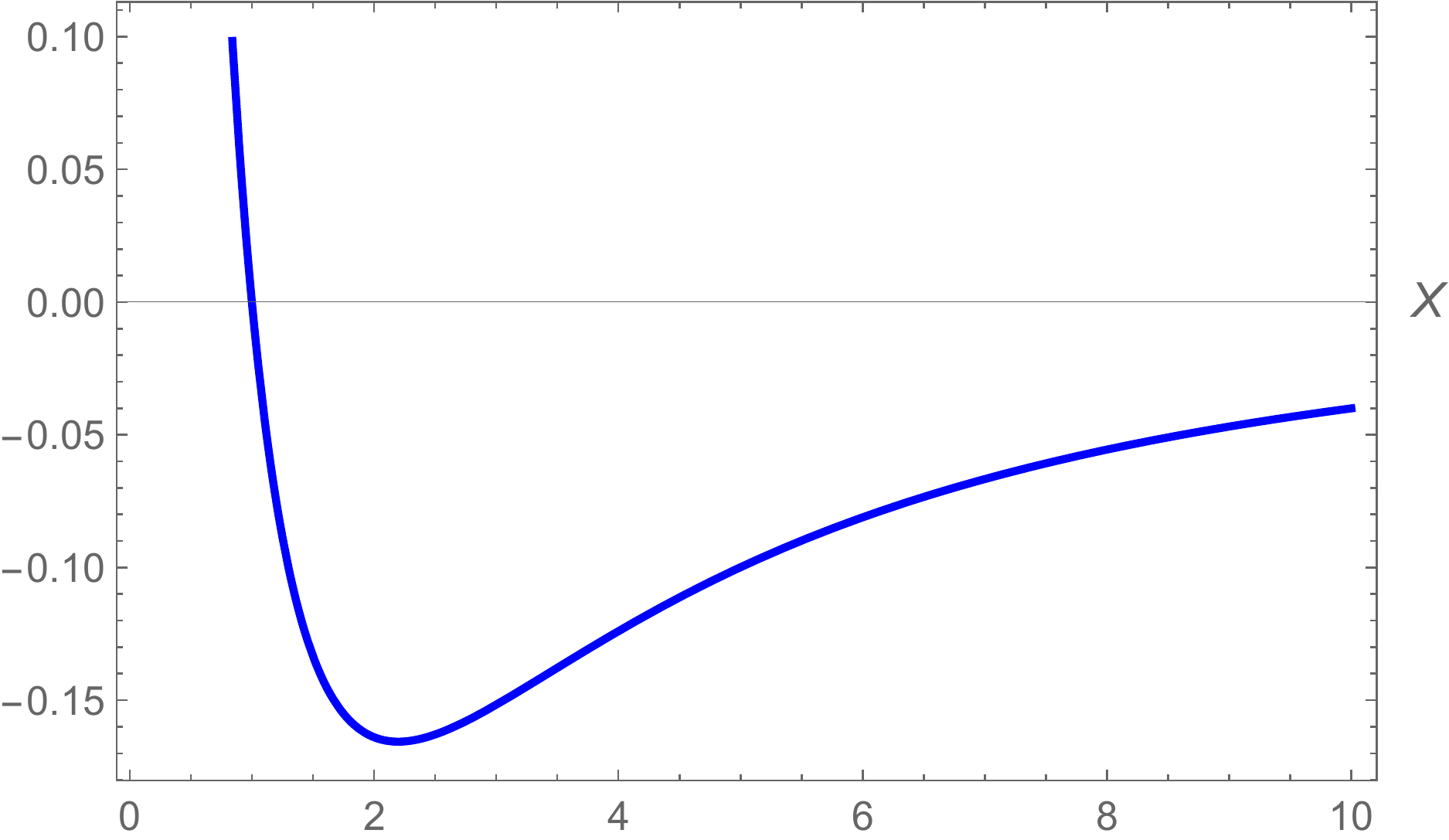}
\caption{ $W_2(X,0)$ vs $X$ for the velocity along the $X$ axis. $S=0.1$, $X$ is in units of $\Lambda$.}
\label{f2}
\end{figure}

\begin{figure}[t ]
\includegraphics[width=7.5cm] {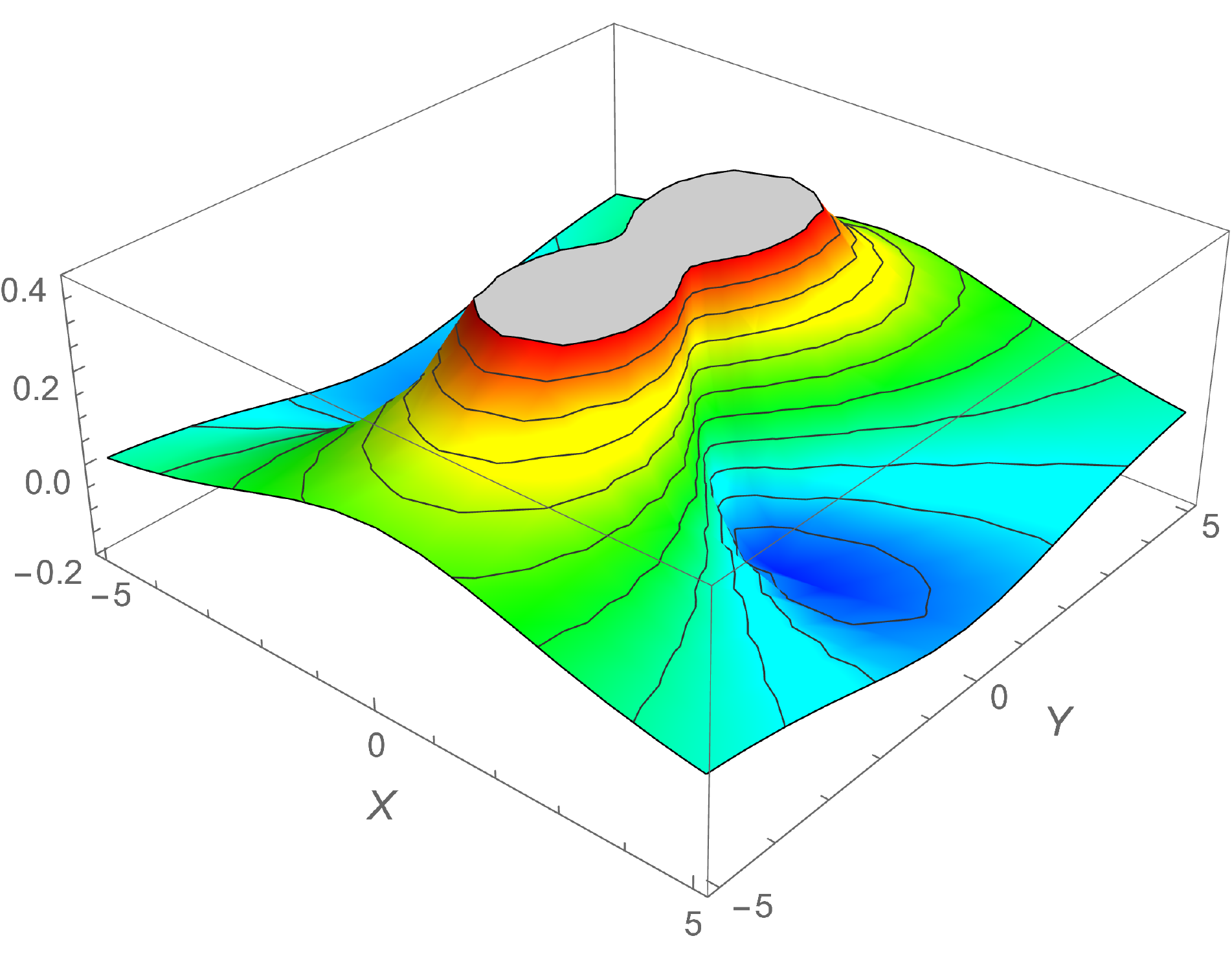}
\caption{3D plot of $W_2(X,Y)$ for the velocity along the $X$ axis.  $(X,Y)=(a_x,a_y)/\Lambda$.}
\label{f3}
\end{figure}

A surprising feature of this plot are the two minima at the $X$ axis situated symmetrically relative to the origin ($X$ is along $\bm v$). One of these minima is shown in Fig.\,\ref{f2} where the graph of $W_2(X,0)$ is plotted to indicate the minimum position at $X_m\approx 2.2$.
To see a clear picture of the dissipation ${\cal W}(\bm a)=W_2(\bm a)+$\,const, we also show the 3D version of the same result in Fig.\,\ref{f3}.

For an arbitrary velocity, one has to keep the term $ik_x v\tau$ in denominators of electric field components (\ref{Exy}). One then obtains
 \begin{eqnarray}
   \frac{{ \cal W}}{{ \cal W}_0}  =  \int \frac{d^2\bm q\, q_x^2 (1+\cos \bm q\bm R)}{  q^2[(1+  q)^2+q_x^2S^2] }   \,.
\label{eq22} 
\end{eqnarray}
The dimensionless parameter  
\begin{eqnarray}
S= v\,\frac{2\pi  \sigma d}{c^2} 
  \label{S}
\end{eqnarray}
is small even for vortex velocities exceeding the speed of sound presently attainable  \cite{Eli,Denis} if one takes for the estimate the conductivity $\sigma$ of normal quasi-particles as equal to the normal state conductivity. 
Unfortunately, there is not much experimental information about the $T$ dependence of $\sigma$. Theoretically, this question is still debated, e.g.  Ref.\,\cite{Andreev} discusses possible strong enhancement of $\sigma$ due to inelastic scattering.

  \begin{figure}[t ]
\includegraphics[width=7cm] {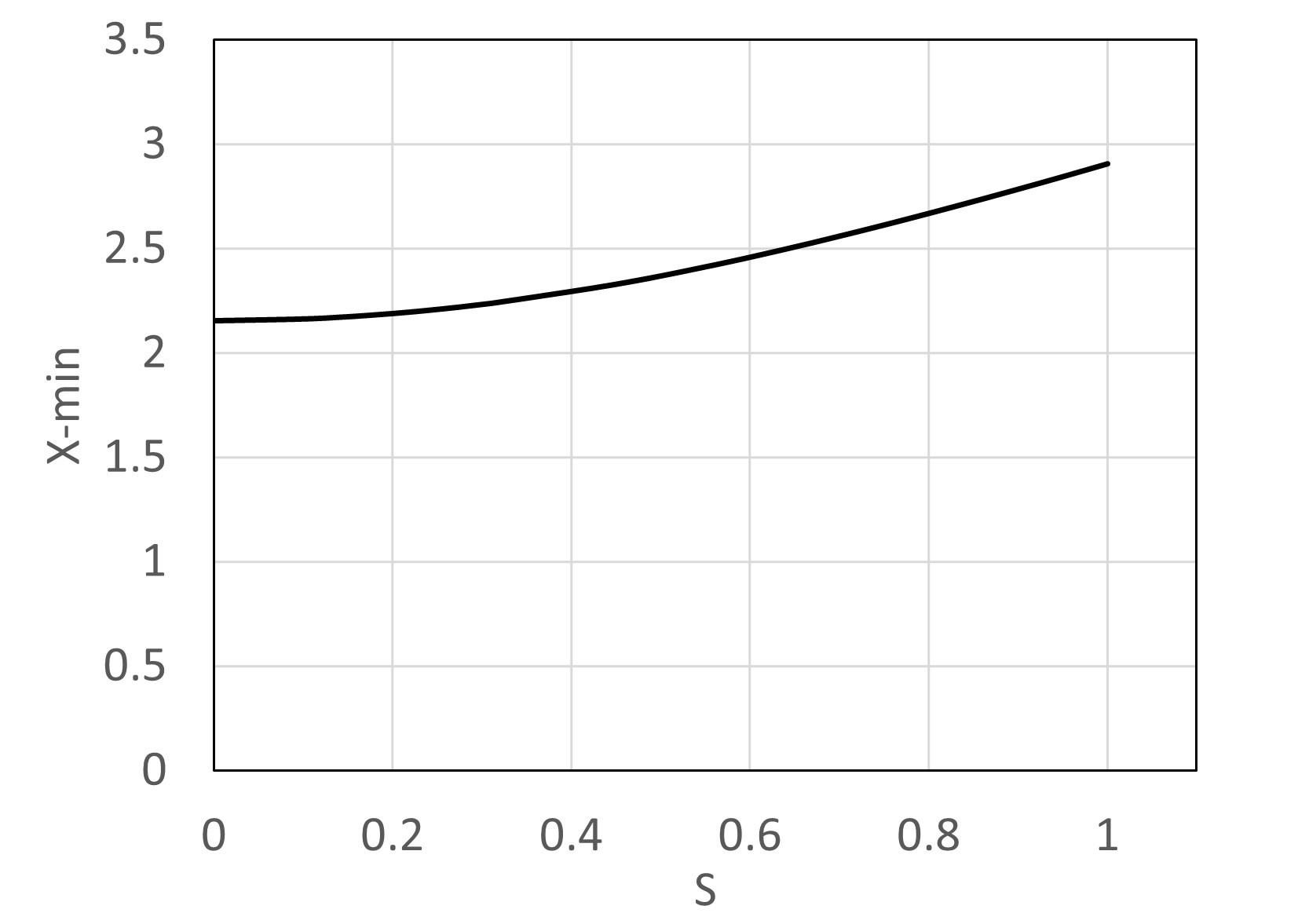}
\caption{The minimum position  $X_m$ vs $S$.   }
\label{f4}
\end{figure}


We   employ the Fast Fourier Transform   to evaluate the integral  (\ref{eq22}). 
The position $X_m$ of the minimum of ${\cal W}(X,0)$  for each S was obtained from the contour plot similar to Fig.\,\ref{f1}, which was
sliced out of the 2D map obtained from the cosine term of Eq.\,(\ref{eq22}) via 2D FFT.
 The result  is shown in Fig.\,\ref{f4}. 
 Hence, for $S < 0.2$, which is the domain of our interest, the minimum is practically in the same place at $X_m = x_m/\Lambda\approx 2.2$.

\section{Discussion}

  
Hence, the dissipation $\cal W$ of two vortices separated by $\bm R=(X,Y)$ depends on the pair orientation relative to the velocity  $\hat{\bm v}$  and on the pair size $R$. 
The  numerically evaluated dissipation   ${\cal W}(X,Y)$ is shown in Fig.\,\ref{f3}. The dissipation power has a minimum if the pair is oriented parallel to $\bm v$ and the vortices are separated by $a_m=R_m\Lambda \approx 2.2\Lambda $. 

  The physical reason for this minimum can be traced to the magnetic structure of a single moving vortex. 
 It was shown in \cite{TDL,norio1} that the  magnetic field is depleted  in front of the moving vortex  and enhanced behind it due to induced currents of normal excitations. If two vortices move so that one follows the other and $\bm a\parallel \bm v$, in the space between them the depletion of the second is compensated by the enhancement due to the leader. The resulting magnetic field variation in this space is weaker than for a single vortex. Then the electric field induced in this intervortex region  $\bm E \propto \partial_t \bm h \propto \bm v\cdot \bm \nabla \bm h $   is suppressed along with the dissipation. Clearly this simple argument does not work if the pair orientation differs from $\bm a\parallel \bm v$.

Moving vortices in Pb films were studied in  \cite{Eli}. The penetration depth of bulk Pb is $\lambda\approx 96\,$nm and the film thickness   $d=75 \,$nm so that  the Pearl length $\Lambda\approx 246\,$nm. Vortices driven across the thin-film bridge by a transport current  are reported to form chains with spacing $a$ depending on the distance from the bridge edge. Since the driving current   decreases with distance $x$ from the edge, the vortex velocity   depends on $x$ as well. The team \cite{Eli} was able to measure both $v(x)$ and $a(x)$.

According to our model, the pair of moving vortices dissipates the least if it is oriented along the velocity and separated by $a_m\approx 2.2\, \Lambda$. One can expect 
the chain of vortices to have a period  of the order   $a_m   $.
 Taking the experimental estimate of $\Lambda $  we obtain $a_m\approx 540\,$nm. In the experiment \cite{Eli} the chain period varies from $\approx 1500\,$nm near the bridge edge to $\approx 600\,$nm (for the set of data with the transprot current $18.9\,$mA). Hence, the order of magnitude provided by our model is correct. In other words, the idea that the chain period is dictated by the minimum of dissipation agrees qualitatively with observations. 
 
  From the data \cite{Eli}, close to the bridge edge the   chain period $a \approx 1.5\,\mu$m and the velocity $v\approx 16$\,km/s, i.e. the ratio $a /v\approx 10^{-10}\,$s. 
On the other hand, the theoretical ratio 
\begin{eqnarray}
\frac{x_m}{v}= 2.2 \,\frac{  \Lambda}{v}  =\frac{4\pi\sigma\lambda^2}{c^2S} \,,  
  \label{xm/v}
\end{eqnarray}
where we replaced the velocity with $S$ according to Eq.\,(\ref{S}).
Taking for $x_m/v$ the experimental ratio $a /v\approx 10^{-10}\,$s and $\lambda\approx 96\,$nm, 
we  estimate the conductivity of normal excitations 
$\sigma\approx (3\times 10^{19}S)\,$s$^{-1}$. With $S\sim 10^{-2}$ this gives the Pb conductivity that again suggests a qualitative relevance of our model.    We note again  that recent theories suggest a higher conductivity of the normal excitations in superconductors than their normal conductivity \cite{Andreev}.





\end{document}